%% file: charm.tex
\documentclass[12pt]{article}
\usepackage{graphicx}

\def\pbnr{}
\def\speaker{E.R. Cazaroto $^a$, V.P. Gon\c{c}alves $^b$ and F.S. Navarra $^a$}
\def\onbehalfof{}
\def\title{Charm production in double parton scattering at the LHC}
\def\affiliation{$^a$ Instituto de F\'{\i}sica, Universidade de S\~{a}o Paulo,
C.P. 66318, \\ 05315-970 S\~{a}o Paulo, SP, Brazil \\
 $^b$ Instituto de F\'{\i}sica e Matem\'atica, Universidade Federal de
Pelotas\\
Caixa Postal 354, CEP 96010-900, Pelotas, RS, Brazil.}
\def\support{}

\input charmmacros.tex

\begin{document}
\begin{titlepage}
\pubblock

\vfill
\Title{\title}
\vfill
\Author{\speaker\SupportedBy{\support}\OnBehalf{\onbehalfof}}
\Address{\affiliation}
\vfill
\begin{Abstract}
We calculate $c\bar{c}c\bar{c}$ and $b\bar{b}b\bar{b}$ production cross section in Double Parton Scattering (DPS) processes at the energies of LHC. In our calculation we include saturation effects in the gluon distribution. For charm production, our results confirm the prediction of a previous work in which saturation effects were neglected, namely that at the energies of LHC charm production in DPS processes becomes comparable with charm production in SPS. We also estimate the production of $c\bar{c}b\bar{b}$ in DPS. Our results indicate that at $\sqrt{s}=14$ TeV half of the total amount of bottom produced in proton-proton collisions comes from this process.
\end{Abstract}
\vfill
\begin{Presented}
\venue
\end{Presented}
\vfill
\end{titlepage}
\def\thefootnote{\fnsymbol{footnote}}
\setcounter{footnote}{0}
%

\section{Introduction}

It is commonly believed that Single Parton Scattering (SPS) exerts the dominant role in hadronic collisions. In this kind of process one gluon from the hadron target scatters with one gluon from the hadron projectile, generating e.g. one heavy quark pair $Q\bar{Q}$ in the final state of the collision. However, the huge density of gluons present in the hadrons in the high energy regime increases the probability of multiple-scatterings to take place in a single proton-proton collision. For example, in one hadron-hadron collision we can have two gluons from the target scattering with two gluons from the projectile, each gluon-gluon fusion happening almost independently. This kind of process is called Double Parton Scattering (DPS). The DPS was recognized and discussed in the 1970's and 1980's \cite{70_80}, but it was soon realized that the cross section for the DPS process is negligible at the energies available at that time. In the last decade several works have been dedicated to the study of the DPS due to its importance in the kinematic regime of the LHC. In particular, in the recent work \cite{Marta_Rafal} (see also \cite{liko_prd86}) the authors obtained the surprising result that at the energies of the LHC the $c\bar{c}c\bar{c}$ production cross section in DPS processes is of the same order of magnitude as the $c\bar{c}$ production cross section in SPS, with the $c\bar{c}c\bar{c}$ production cross section in SPS being strongly suppressed. In Ref. \cite{dps_nos} we confirmed that this result remains valid when saturation effects in the gluon distribution are taken into account. In this work we review some of the results that we obtained in Ref. \cite{dps_nos} and add a new comparison of the DPS cross sections for three final states, namely $c\bar{c}c\bar{c}$, $b\bar{b}b\bar{b}$ and $c\bar{c}b\bar{b}$.


\section{The heavy quark production in DPS}
\label{sig_dps}

In a rigorous treatment one should consider non-trivial correlations between the gluons involved in the DPS process. Several works have been dedicated to study different kinds of correlations (See, e.g., Ref. \cite{wouter1}), however it has been shown to be very difficult to make a precise estimate of their magnitude. As a consequence, in practical calculations, the authors disregard any correlation between the gluons and factorize the DPS cross section as a product of two completely independent SPS cross sections. With this assumption the DPS cross section for $Q_1\bar{Q}_1Q_2\bar{Q}_2$ production is given by (See, e.g., Ref. \cite{diehl_jhep}):
\begin{eqnarray}
\sigma_{h_1 h_2 \rightarrow Q_1\bar{Q}_1Q_2\bar{Q}_2}^{DPS} = \left( \frac{m}{2} \right) \frac{\sigma^{SPS}_{h_1 h_2 \rightarrow Q_1\bar{Q}_1} 
\sigma^{SPS}_{h_1 h_2 \rightarrow Q_2\bar{Q}_2}}{\sigma_{eff}} \,\,,
\label{dps_fac}
\end{eqnarray}
where $m/2$ is a combinatorial factor which accounts for the indistinguishable or distinguishable final states, i.e., $m=1$ when $Q_1=Q_2$ and $m=2$ when $Q_1 \neq Q_2$. The free parameter $\sigma_{eff}$ was determined by CDF collaboration through a fit to experimental data and it was estimated to be $\sigma _{eff} = (14.5 \pm 1.7_{-2.3}^{+1.7})$ mb \cite{cdf}. In what follows we will use Eq. (\ref{dps_fac}) to estimate the $Q_1\bar{Q}_1Q_2\bar{Q}_2$ production cross section in DPS processes with $\sigma _{eff} = 15$ mb.

Saturation effects in the gluon distribution are naturally described in the color dipole formalism (See \cite{hqp_nos} and references therein). This formalism makes use of the fact that, before interacting with the gluon target, the gluon emitted by the projectile fluctuates into a color octet pair $Q\bar{Q}$. In this approach we have to choose a model for the color dipole-target cross section. This model contains all the information about the QCD dynamics at high energies, including the presence of non-linear corrections that leads to the gluon saturation. We will use two saturation models and compare their results. The first model is a numerical solution of the Balitsky-Kovchegov (BK) equation \cite{bkrunning}, which will be labeled as ``rcBK''. The second model is the one proposed by Golec - Biernat and Wusthoff in Ref. \cite{gbw}, which will be labeled as ``GBW''. 
Our motivation to use the GBW model is that it allows us to easily  obtain its linear limit. Consequently, by using the two models, ``GBW'' and ``GBW Linear'', in the calculations we can compare its results and quantify the contribution of the saturation effects in the observable under analysis.

\section{Results and discussions}
\label{results}

\begin{figure}[htb]
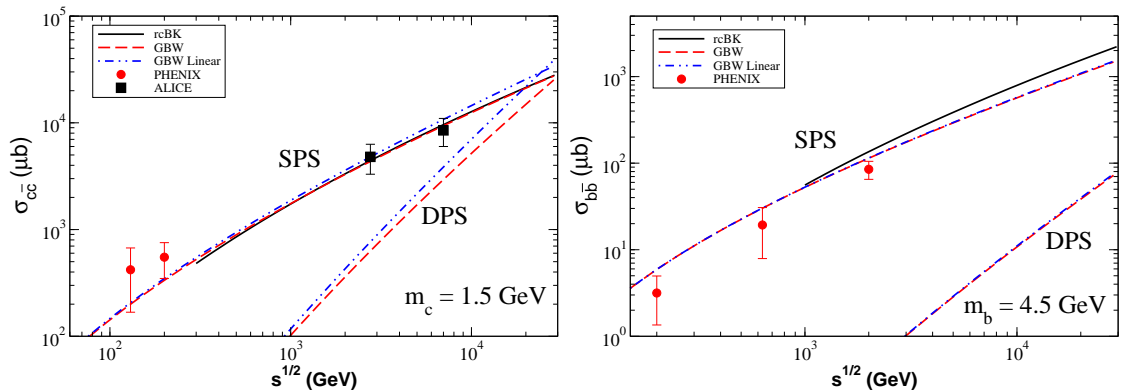

\center
\begin{tabular}{cc}
\includegraphics[scale=0.29]{sig_tot_e_dps_charm.eps}        
\includegraphics[scale=0.29]{sig_tot_e_dps_bottom.eps}        
\end{tabular}
\caption{Charm (left) and bottom (right) production cross sections in Single Parton Scattering (SPS) and Double Parton Scattering (DPS) as a function of the c.m.s. energy ($\sqrt{s}$). Data points from PHENIX \cite{phenix} (circles) and  from  ALICE \cite{ALICE_pp} (squares).}
\label{fig:2}
\end{figure}

Since the models that we are using have their parameters already fixed, in our calculation the only free parameter is the heavy quark mass. So the first step is to constrain the heavy quark mass by adjusting its value to fit the available experimental data. To describe at the same time the data from PHENIX \cite{phenix} and the recent data from ALICE \cite{ALICE_pp} on charm production we fixed the charm mass in $m_c = 1.5$ GeV (Fig. \ref{fig:2} - Left). It is interesting to observe how the ALICE data were able to reduce the freedom of choice of $m_c$. Before their appearance, the existing data could be fitted with values of $m_c$ in the range $1.2$ GeV $\leq m_c \leq 1.5$ GeV, as shown in \cite{hqp_nos}. Now, the lowest value ($m_c=1.2$ GeV) is excluded. For bottom production we do not have available high energy data from LHC, so we used $m_b = 4.5$ GeV for the bottom mass and compared our results with the data from PHENIX (Fig. \ref{fig:2} - Right).
In Fig. \ref{fig:2} we see that the DPS cross section for charm production becomes comparable with the SPS one at the energies of LHC. This result was first obtained in Ref. \cite{Marta_Rafal}, where saturation effects were neglected. Now we are confirming that this result remains valid when saturation is taken into account. For bottom production we see that the DPS cross section is negligible when compared with the SPS one in the whole range of energy. Note that the rcBK prediction for charm production practically coincides with the GBW one. So from now on we will use only the GBW model to make the predictions of saturation physics.

\begin{figure}[htb]
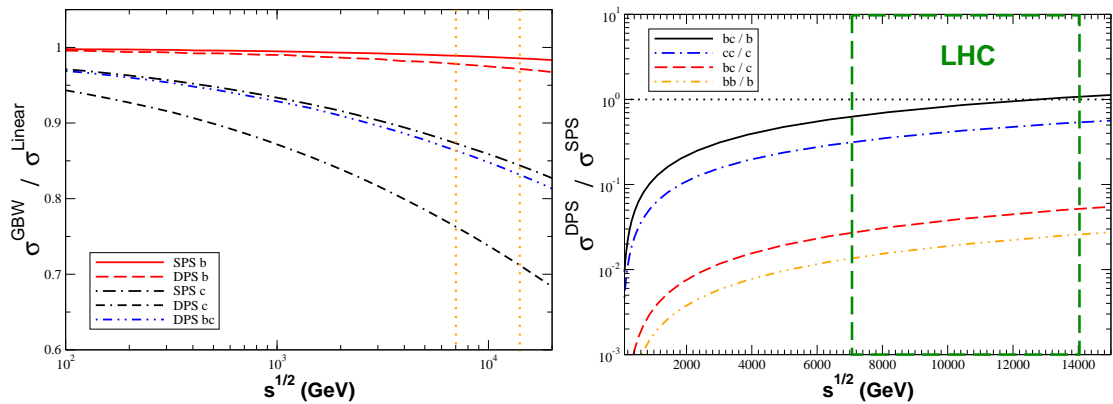

\center
\begin{tabular}{cc}
\includegraphics[scale=0.3]{razao_gbw_por_linear.eps}
\includegraphics[scale=0.3]{razoes_dps_por_sps.eps}
\end{tabular}
\caption{Left: The ratio $\, \sigma^{GBW} / \sigma^{GBW Linear} \,$ in SPS and DPS processes as a function of the c.m.s. energy ($\sqrt{s}$); Right: The ratio $\sigma ^{DPS}/\sigma ^{SPS}$ as a function of the c.m.s. energy ($\sqrt{s}$).}
\label{fig:3}
\end{figure}

In order to estimate more precisely the magnitude of saturation effects in SPS as well as in DPS processes we plotted in Fig. \ref{fig:3} - Left the ratio $\, \sigma^{GBW} / \sigma^{GBW Linear} \,$. As we can see the production of bottom in SPS as well as in DPS is practically insensitive to saturation effects, the ratio being approximately 1 in the whole range of energy. On the other hand, the charm production is very sensitive to saturation effects in the considered range of energy. In particular, at $\sqrt{s} = 14$ TeV the charm production in SPS  is decreased by $\approx 15\%$ whereas the charm production in DPS is decreased by $\approx 28\%$. We also considered the production of $c\bar{c}b\bar{b}$ in DPS process (``DPS bc'' in the legend). We can see that the corresponding curve is very close to the curve of charm production in SPS process. This is a consequence of the fact that the sensitiveness of this process to saturation effects comes from the charm sector, whose production is much more sensitive to saturation effects than the production of bottom. 

We have also investigated the importance of DPS processes when compared with SPS ones. In Fig. \ref{fig:3} - Right we plotted the ratio $\sigma ^{DPS}/\sigma ^{SPS}$. The legend `` bc / b '' means that we are taking $b\bar{b}c\bar{c}$ production in DPS divided by $b\bar{b}$ production in SPS, and so on. The larger ratio (full black line) is the one corresponding to `` bc / b ''. At $\sqrt{s} = 14$ TeV this ratio becomes $\approx 1$, what means that the DPS cross section of $b\bar{b}c\bar{c}$ production becomes equal the SPS cross section of $b\bar{b}$ production. In other words, half of the total amount of bottom produced at the LHC in $\sqrt{s} = 14$ TeV will come from the DPS channel. The second curve in magnitude is the one labeled as `` cc / c ''. This curve reaches $\approx 0.6$ in $\sqrt{s} = 14$ TeV, which means that $\approx 1/3$ of the total amount of charm that will be produced at the LHC in this energy will come from the DPS channel.

\begin{figure}[htb]
\center
\begin{tabular}{cc}
\includegraphics[scale=0.45]{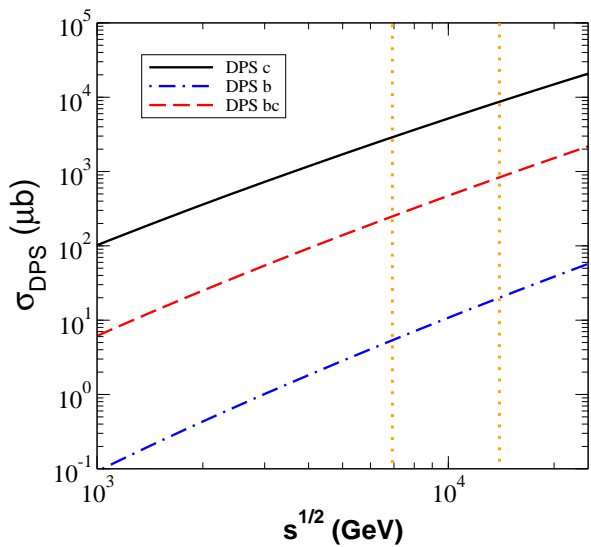}
\end{tabular}
\caption{DPS production cross section for three final states: $c\bar{c}c\bar{c}$, $b\bar{b}b\bar{b}$ and $c\bar{c}b\bar{b}$.}
\label{fig:4}
\end{figure}

Finally, in Fig. (\ref{fig:4}) we show our results for the DPS production cross section for three final states, namely $c\bar{c}c\bar{c}$, $b\bar{b}b\bar{b}$ and $c\bar{c}b\bar{b}$. We can see that all the curves follow the same behavior. This is a consequence of the fact that the only free parameter in our calculation is the quark mass. So, by increasing the mass of the final state we just lower the curve. 

The DPS approach has been successfully applied to the study of several observables, as discussed, for example, in \cite{diehl_jhep}. However it grows too fast with energy $\sqrt{s}$. One of our motivations to include saturation effects in DPS was to tame this fast rise of the cross section. However the observed reduction of the growth of $\sigma _{DPS}$ is not very pronounced and at higher energies some unitarization procedure will be required.

\Acknowledgements
This work was  partially financed by the Brazilian funding agencies FAPESP, CNPq, CAPES and FAPERGS.

\end{document}

%% file: charmmacros.tex
\newcommand\pubnumber{\pbnr}
\newcommand\pubdate{\today}
\def\Title#1{\begin{center} {\Large #1 } \end{center}}
\def\Author#1{\begin{center}{ \sc #1} \end{center}}

\newcommand{\OnBehalf}[1]{\sbox0{#1}\ifdim\wd0=0pt
        {}
	\else
	{\\on behalf of #1}
	\fi}
\newcommand{\SupportedBy}[1]{\sbox0{#1}\ifdim\wd0=0pt
        {}
	\else
	{\footnote{#1}}
	\fi}
\def\Address#1{\begin{center}{ \it #1} \end{center}}

\newcommand\pubblock{\includegraphics[width=5cm]{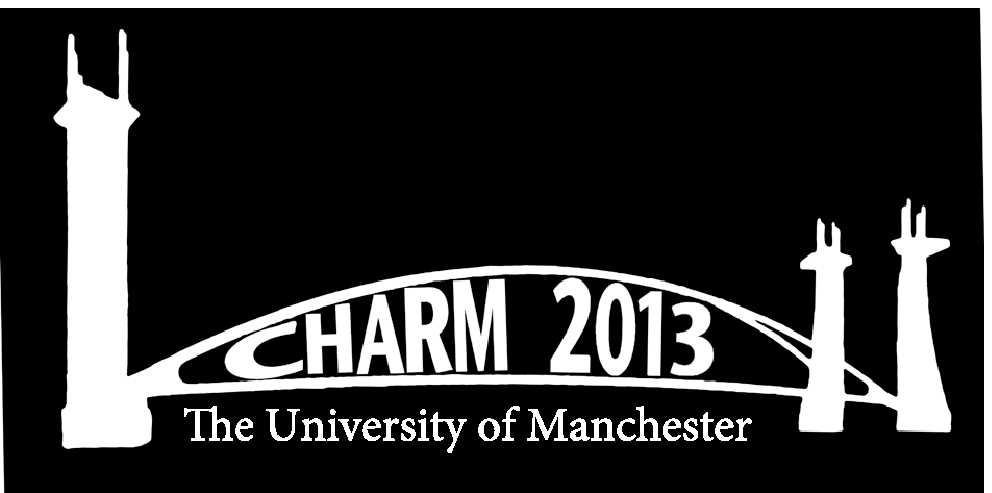}\hfill{\begin{tabular}{l} \pubnumber\\
         \pubdate  \end{tabular}}}
\newenvironment{Abstract}{\begin{quotation}  }{\end{quotation}}
\newenvironment{Presented}{\begin{quotation} \begin{center} 
             PRESENTED AT\end{center}\bigskip 
      \begin{center}\begin{large}}{\end{large}\end{center} \end{quotation}}
\def\Acknowledgements{\bigskip  \bigskip \begin{center} \begin{large}
             \bf ACKNOWLEDGEMENTS \end{large}\end{center}}
\def\venue{The 6$^{th}$ International Workshop on Charm Physics\\
(CHARM 2013)\\
Manchester, UK,  31 August -- 4 September, 2013}

\def\beq{\begin{equation}}
\def\eeq#1{\label{#1}\end{equation}}
\def\eeqn{\end{equation}}

\def\beqa{\begin{eqnarray}}
\def\eeqa#1{\label{#1}\end{eqnarray}}
\def\eeqan{\end{eqnarray}}

\let\bar=\overbar

\def\Dslash{\not{\hbox{\kern-4pt $D$}}}
\def\dslash{\not{\hbox{\kern-2pt $\del$}}}

\def\msb{{\bar{\ssstyle M \kern -1pt S}}}

